\documentclass[11pt]{article}

\usepackage[top=70pt,bottom=80pt,left=95pt,right=95pt]{geometry}

\usepackage{amsmath, amssymb, amsfonts, latexsym, euscript, mathalfa, mathrsfs}
\usepackage{graphicx, color, epsfig, xcolor}
\usepackage{cite}
\usepackage{empheq}
\usepackage{pictexwd,dcpic}

\usepackage{setspace, sectsty}

\usepackage{braket, cancel}

\usepackage[debug,pageanchor=false]{hyperref}
\definecolor{link}{rgb}{.8,.15,.1}
\hypersetup{colorlinks=true,linkcolor=link,citecolor=link,urlcolor=link,linktocpage}

\makeatletter
\@addtoreset{equation}{section}
\makeatother

\newlength{\sswidth}

\newcommand{\nn}{\nonumber}

\def\be{\begin{equation}}
\def\ee{\end{equation}}
\newcommand{\eq}[1]{\begin{equation}\begin{split}#1\end{split}\end{equation}}

\newcommand{\sub}[1]{\begin{subequations}\begin{align}#1\end{align}\end{subequations}}

\def\e{\epsilon}
\def\g{\gamma}

\def\m{m}
\def\o{\omega}
\def\O{\Omega}
\renewcommand{\t}{\theta}

\newcommand{\vol}{\mathrm{vol}}

\newcommand{\I}{\text{Im}}
\newcommand{\R}{\text{Re}}
\newcommand{\p}{\partial}

\begin{document}

\begin{titlepage}

\begin{flushright} \small

\end{flushright}

\begin{center}

\noindent

{\Large \bf On AdS$_3$ solutions of Type IIB}

\bigskip\medskip

Achilleas Passias$^1$ and Dani\"{e}l Prins$^{2,3}$ \\

\bigskip\medskip
{\small

$^1$ D\'{e}partement de Physique, \'{E}cole Normale Sup\'{e}rieure, Universit\'{e} PSL, CNRS, \\
24 Rue Lhomond, 75005 Paris, France
\\	
\vspace{.3cm}
$^2$Institut de Physique Th\'{e}orique, Universit\'{e} Paris Saclay, CNRS, CEA, \\
F-91191 Gif-sur-Yvette, France
\\	
\vspace{.3cm}
$^3$Dipartimento di Fisica, Universit\`a di Milano--Bicocca, \\ Piazza della Scienza 3, I-20126 Milano, Italy \\ and \\ INFN, sezione di Milano--Bicocca
}


\vskip .9cm 
     	{\bf Abstract }
\vskip .1in
\end{center}

We study $\mathcal{N}=1$ supersymmetric AdS$_3 \times M_7$ backgrounds of Type IIB supergravity, with non-vanishing axio-dilaton, three-form and five-form fluxes, and a ``strict'' $SU(3)$-structure on $M_7$.
We derive the necessary and sufficient conditions for supersymmetry as a set of constraints on the torsion classes of the $SU(3)$-structure.
Given an Ansatz for the three-form fluxes, the problem of also solving the equations of motion involves a ``master equation'', which generalizes
ones that have previously appeared in the literature.

\vfill

\eject

\end{titlepage}

\tableofcontents

\section{Introduction}

Recently, there has been renewed interest in supersymmetric AdS$_3 \times M_7$ backgrounds of Type IIB supergravity dual to $(0,2)$ superconformal field theories (SCFTs) in two dimensions \cite{Couzens:2018wnk, Gauntlett:2018dpc}. In particular, the authors of \cite{Couzens:2018wnk, Gauntlett:2018dpc} studied such backgrounds with only five-form flux \cite{Kim:2005ez}, and showed the existence of the geometric dual of $c$-extremization in two-dimensional $(0,2)$ SCFTs \cite{Benini:2012cz, Benini:2013cda}.

Motivated by the expectation that a geometric dual of $c$-extremization should exist for more general backgrounds than the ones considered in \cite{Couzens:2018wnk, Gauntlett:2018dpc}, we aim to provide, as a first step, a systematic classification of supersymmetric AdS$_3 \times M_7$ backgrounds of Type IIB supergravity.\footnote{A similar expectation for the geometric dual of $a$-maximization in four dimensions \cite{Intriligator:2003jj} was explored in \cite{Gabella:2010cy} using generalized geometry.} Such a classification was initiated by the author of \cite{Kim:2005ez}, with the backgrounds mentioned in the previous paragraph (see also \cite{Gauntlett:2007ts}). In \cite{Donos:2008ug}, this class was extended to also admit a three-form flux satisfying certain conditions, whereas in \cite{Couzens:2017nnr} instead a varying axio-dilaton was included. In this note we extend this classification program further, by allowing for both varying axio-dilaton and three-form fluxes. 
Although generically we classify supersymmetric backgrounds that are dual to $(0,1)$ SCFTs\footnote{See \cite{Eberhardt:2017uup} for backgrounds with pure NSNS flux.} for which no principle of $c$-extremization exists, as we discuss below, our study does lead to solutions dual to $(0,2)$ SCFTs.
We restrict to the case that $M_7$ is equipped with a ``strict'' $SU(3)$-structure, which is equivalent to requiring that the two Majorana supersymmetry parameters on $M_7$ are orthogonal. Our classification includes as special cases the ones by
\cite{Kim:2005ez, Couzens:2018wnk}.
The necessary and sufficient conditions for supersymmetry are phrased as
restrictions on the torsion classes of the $SU(3)$-structure,
which in seven dimensions is determined by a real one-form $v$, a real two-form $J$, and a complex decomposable three-form $\Omega$.
The vector dual to $v$ foliates $M_7$, and we find that the transverse six-dimensional space $M_6$ is conformally symplectic.

\begin{figure}
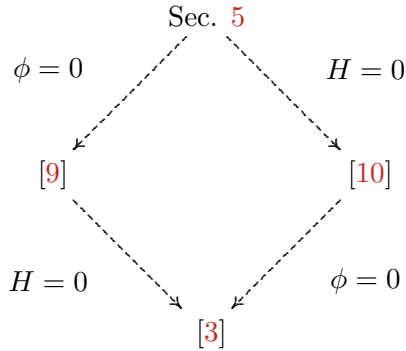

\label{classes}
\begin{align*}
\begindc{\commdiag}[200]
\obj(+0,+3)[a]{Sec. \ref{newclass}}
\obj(-3, 0)[b]{\cite{Donos:2008ug}}
\obj(+3, 0)[c]{\cite{Couzens:2017nnr}}
\obj(-0,-3)[d]{\cite{Kim:2005ez}}
\obj(+3,-2)[e]{$\phi = 0$}
\obj(-3,-2)[e]{$H = 0$}
\obj(+3,+2)[e]{$H = 0$}
\obj(-3,+2)[e]{$\phi = 0$}
\mor{a}{b}{}[+1,1]
\mor{a}{c}{}[+1,1]
\mor{b}{d}{}[+1,1]
\mor{c}{d}{}[+1,1]
\enddc
\end{align*}
\caption{Depiction of the relation between classes of solutions. $\phi$ is the dilaton, and $H$ the
NSNS flux.}
\end{figure}

On AdS$_3 \times M_7$, a solution to the supersymmetry equations also solves the equations of motion if and only if the Bianchi identities are satisfied by the fluxes (see for example \cite{Prins:2013wza}).
By making an Ansatz for the three-form fluxes in our solution to the supersymmetry equations,
we reduce the problem of finding a solution to the Bianchi identities, and hence the equations of motion, to two conditions:
a ``master equation'' \eqref{master}, which is a partial differential equation for the conformally K\"{a}hler metric on $M_6$,
and existence of a primitive (1,2)-form satisfying \eqref{constraint}. Furthermore, supersymmetry is
enhanced to $\mathcal{N} = 2$.
Similar master equations (and solutions thereof) associated with Bianchi identities appeared in \cite{Kim:2005ez, Donos:2008ug, Couzens:2017nnr}, and the one presented here reduces to the ones
of \cite{Kim:2005ez, Donos:2008ug, Couzens:2017nnr} in the appropriate limits.\footnote{See \cite{Gauntlett:2006ns, Benini:2015bwz} for more solutions dual to two-dimensional $(0,2)$ SCFTs.}
The relation of these classes of solutions, and
the corresponding master equations is depicted in Figure
\ref{classes}. Solutions to the aforementioned conditions, as
well as more general Ans\"{a}tze will be reported in future work.

The rest of this note is organized as follows.
In section \ref{susy}, we present the supersymmetry
equations as a set of equations involving a
pair of polyforms on $M_7$. In section \ref{gstructures}, we introduce an $SU(3)$-structure
in seven dimensions, and parameterize the polyforms
in terms of it. In section \ref{classification},
we derive a set of necessary and sufficient
conditions for supersymmetry as restrictions
on the torsion classes of the $SU(3)$-structure,
and also give expressions for the fluxes in terms of the latter. A summary at the end of this section
is included. Section \ref{newclass} presents
a class of solutions to the equations of motion following an Ansatz, as described earlier. Our conventions and certain technical details are included in the appendix.

\section{Supersymmetry equations} 
\label{susy}

We start with a general bosonic background of Type IIB
supergravity invariant under $SO(2,2)$. The
ten-dimensional metric is a warped product of a metric on
AdS$_3$ and a metric on a seven-dimensional Riemannian manifold
$M_7$:
\eq{
g_{10} = e^{2A} g_{{\rm AdS}_3} + g_{M_7}~,
}
where $A$ is a function on $M_7$.\footnote{We work in string frame.} Conforming to the $SO(2,2)$ symmetry,
the NSNS field-strength $H_{10d}$ and the RR field-strengths $F_{10d}$, with $F_{10d}$
denoting their sum in the democratic formulation, are decomposed as
\begin{equation}
H_{10d} = \varkappa e^{3A} \mathrm{vol}_{{\rm AdS}_3} + H ~,
\qquad
F_{10d} = e^{3A} \mathrm{vol}_{{\rm AdS}_3} \wedge \star_7 \lambda(F) + F ~.
\end{equation}
The magnetic fluxes $H$ and $F = \sum_{p=1,3,5,7} F_p$, are forms on $M_7$. The operator $\lambda$ acts on a $p$-form
$F_p$ as $\lambda(F_p) = (-1)^{\lfloor p/2 \rfloor} F_p$. The RR field-strengths are subject to $d_{H_{10d}}F_{10d}=0$, which decomposes as
\begin{equation}\label{eq:dec-Bianchi}
d_H (e^{3A} \star_7 \lambda(F)) + \varkappa F = 0~,
\qquad
d_H F = 0 ~,
\end{equation}
where $d_H \equiv d - H \wedge$. We will refer to the first set of equations as equations of motion for $F$,
and to the second one as the Bianchi identities.

In order to study the restrictions imposed by supersymmetry on the above bosonic background, we decompose the supersymmetry parameters of Type IIB supergravity, $\e_1$ and $\e_2$ under
Spin$(1,2) \times$ Spin$(7) \subset$ Spin$(1,9)$:\footnote{For the decomposition of the Clifford algebra see the appendix.}
\eq{
\e_1 = \zeta \otimes \chi_1 \otimes \left(\begin{array}{c} 1 \\ -i \end{array}\right) ~,
\qquad
\e_2 = \zeta \otimes \chi_2 \otimes \left(\begin{array}{c} 1 \\ - i \end{array}\right) ~.
}
Here, $\chi_{1}$ and $\chi_{2}$ are Majorana Spin$(7)$ spinors; $\zeta$ is a
Majorana Spin$(1,2)$ spinor satisfying the Killing equation:
\begin{equation}
\nabla_\mu \zeta = \frac{1}{2} \m \gamma_\mu \zeta ~,
\end{equation}
where the real constant parameter $m$ is related to the AdS$_3$ radius $L_{{\rm AdS}_3}$ as $L^2_{{\rm AdS}_3} = 1/m^2$.
The above decomposition follows the requirement for $\mathcal{N}=1$ supersymmetry.

The necessary and sufficient conditions for preserving $\mathcal{N}=1$ supersymmetry can be derived following
the derivation for Type IIA supergravity in the appendix of \cite{Dibitetto:2018ftj}, with straightforward modifications.
They are expressed in terms of bispinors $\psi_\pm$ defined by
\eq{\label{polydef}
\chi_1 \otimes \chi_2^t \equiv \psi_+ + i \psi_- ~.
}
Following the Fierz expansion of $\chi_1 \otimes \chi_2^t$, and application of the Clifford map which maps anti-symmetric products of gamma matrices to forms, $\psi_+$/$\psi_-$ become polyforms on $M_7$, of even/odd degree.

The supersymmetry restrictions take the form of the following system of
equations:
\begin{subequations}
\begin{align}
2 \m c_- &= - c_+ \varkappa ~, \\
\label{eq:c-F} d_H(e^{A-\phi} \psi_+) &= \frac{1}{16} c_- F ~, \\
\label{eq:c+F} d_H(e^{2A-\phi} \psi_-) + 2 \m e^{A-\phi} \psi_+ &= \frac{1}{16} c_+ e^{3A} \star_7 \lambda(F) ~, \\
(\psi_+,F)_7 &= \frac \m2 e^{-\phi} \mathrm{vol}_7 ~.
\end{align}
\end{subequations}
Here $c_\pm$ are constants defined by the norms of $\chi_1$ and $\chi_2$:
\begin{equation}
	c_\pm \equiv e^{\mp A}(||\chi_1||^2 \pm ||\chi_2||^2) ~.
\end{equation}
Furthermore,  $(\psi_+,F)_7 \equiv (\psi_+ \wedge \lambda(F))_7$, with
$(\cdot)_7$ denoting the restriction to the seven-form component.

In this work we will consider backgrounds with zero electric component for $H_{10d}$ i.e.\ $\varkappa = 0$, since an electric component can be set to zero by applying an $SL(2,\mathbb{R})$ duality transformation.\footnote{We thank N.\ Macpherson for pointing this out.} Supersymmetry then dictates $c_- = 0$, or
equivalently $||\chi_1||^2 = ||\chi_2||^2$. The system of supersymmetry equations thus becomes:
\begin{subequations}\label{SUSY}
\begin{align}
d_H(e^{A-\phi} \psi_+) &= 0 ~, \label{SUSYa}\\
d_H(e^{2A-\phi} \psi_-) + 2 \m e^{A-\phi} \psi_+ &= \frac{1}{8} e^{3A} \star_7 \lambda(F) ~, \label{SUSYb}\\
(\psi_+, F)_7 &= \frac \m2 e^{-\phi} \mathrm{vol}_7 ~. \label{SUSYc}
\end{align}
\end{subequations}
Without loss of generality we have set $c_+ = 2$ i.e.\ $||\chi_1||^2 = ||\chi_2||^2 = e^A$.

\section{Supersymmetry and $G$-structures}\label{gstructures}
A nowhere-vanishing Majorana spinor $\chi$ on $M_7$ defines a $G_2$-structure for $TM_7$. A pair of nowhere-vanishing Majorana spinors $\chi_1$, $\chi_2$ define a $G_2 \times G_2$-structure on the generalized tangent bundle $TM_7 \oplus T^*M_7$.
If $\chi_{1}$, $\chi_{2}$ are parallel, the $G_2 \times G_2$-structure reduces to a $G_2$-structure, whereas if $\chi_{1}$, $\chi_{2}$ are orthogonal it reduces to a ``strict'' $SU(3)$-structure. This can be illustrated by the decomposition of $\chi_2$ in terms
of $\chi_1$ (taking $\chi_1, \chi_2$ to be of equal norm):
\eq{\label{chi_rel}
\chi_2 =  \sin \t  \chi_1 - i \cos \t v_m \g^m \chi_1 ~,
}
where $v$ is a real one-form with $||v||= 1$, and $\t \in [0,\pi/2]$.
As $\theta$ varies from $0$ to $\pi/2$, the $G_2 \times G_2$-structure
varies from a ``strict'' $SU(3)$-structure, to an ``intermediate'' $SU(3)$-structure, to a $G_2$-structure.
In this work we will consider the first case, i.e.\ $\theta = 0$.

An $SU(3)$-structure on $M_7$ is defined by a real one-form $v$, a real two-form
$J$, and a complex decomposable three-form $\Omega$, all nowhere-vanishing,  satisfying\footnote{$X \lrcorner \o_{(k)} \equiv \frac{1}{k-1!} X^n \o_{n m_1 ... m_{k-1}} dx^{m_1} \wedge...\wedge dx^{m_{k-1}}.$}
\begin{align}
v \lrcorner J = v \lrcorner \Omega = 0~, \quad
\Omega \wedge J = 0 ~, \quad
\frac{i}{8}\Omega \wedge \overline\Omega = \frac{1}{3!} J \wedge J \wedge J~.
\end{align}
These forms can be expressed as bilinears in terms of the spinors $(\chi_1, \chi_2)$; see appendix A for our conventions.
The one-form $v$ gives a foliation of $M_7$ with leaves $M_6$; accordingly, we define the volume form as $\vol_7 \equiv \frac{1}{3!} v \wedge J \wedge J \wedge J$ and locally decompose the metric on $M_7$ as
\eq{
g_{M_7} = v \otimes v + g_{M_6}~.
}

Existence of an $SU(3)$-structure ensures that all forms on $M_7$ decompose into irreducible representations of $SU(3)$. In particular,
the local $k$-forms with no component along $v$ can be decomposed into primitive $(p,q)$-forms.\footnote{A primitive $k$-form $\o^{(k)}$ satisfies $J \lrcorner \o^{(k)} = 0$ for $k=2,3$, whereas $k$-forms with $k=0,1$ are primitive by definition. The $(p,q)$ decomposition of $k$-form $\o $ is defined by
\eq{
\o^{(p,q)}_{m_1...m_k} &= \frac{k!}{p!q!}
 \left(\Pi^+\right)_{[m_1}^{\phantom{m}n_1} ... \left(\Pi^+\right)_{m_p]}^{\phantom{m}n_p}
 \left(\Pi^-\right)_{[m_{p+1}}^{\phantom{m}n_{p+1}} ... \left(\Pi^-\right)_{m_k]}^{\phantom{m}n_k}\o_{n_1...n_k}~,\\
 \left(\Pi^{\pm}\right)_m^{\phantom{m}n} &= \frac12 \left( \delta_m^n \mp i J_m^{\phantom{m}n} - v_m v^n \right)~.
}
}

We may also apply this decomposition to the exterior derivatives of the $SU(3)$-structure $\{v,\, J,\, \O\}$ itself. Doing so, we find a parameterization in terms of torsion classes. These constitute the components of the intrinsic torsion of the $SU(3)$-structure expressed in irreducible representations of $SU(3)$. Specifically, we have (see for example \cite{Cassani:2012pj})
\eq{
dv &= R J + T_1 + \R (\overline{V_1} \lrcorner \O) + v \wedge W_0~,  \\
dJ &= \frac32 \I ( \overline{W}_1 \O) + W_3 + W_4 \wedge J
+ v \wedge \left( \frac23 \R E J  + T_2 + \R (\overline{V_2} \lrcorner \O) \right)~,  \\
d \O &= W_1 J \wedge J + W_2 \wedge J + \overline{W_5} \wedge \O
+ v \wedge \left(E \O - 2 V_2 \wedge J + S \right)~.
}
The real scalar $R$ and the complex scalars $E$ and $W_1$ transform in the
$\bf 1$ representation of $SU(3)$. The complex $(1,0)$-forms $V_1$, $V_2$ and $W_5$ transform in the $\bf 3$, and the real one-forms $W_0$ and $W_4$ in the $\bf 3 + \overline{\bf 3}$. The real primitive (1,1)-forms $T_1$ and $T_2$, and the complex primitive (1,1)-form $W_2$ transform in the $\bf 8$. Finally, the real primitive $(2,1)+(1,2)$-form $W_3$ transforms in the $\bf 6 + \overline{\bf 6}$, and the complex primitive $(2, 1)$-form $S$ in the $\bf 6$.

In order to solve the supersymmetry equations, we  parameterize the polyforms $\psi_\pm$ as defined in \eqref{polydef} in terms
of the $SU(3)$-structure data. Making use of \eqref{chi_rel}, \eqref{etadef}, \eqref{bilinears} we find that in the general case,
\eq{
\psi_+^{G_2 \times G_2} &=  \frac{1}{8}  e^A \left[  \I(e^{i \t} e^{iJ}) + v\wedge \R ( e^{i \t} \O ) \right] ~, \\
\psi_-^{G_2 \times G_2} &=  \frac{1}{8}  e^A \left[ v \wedge \R(e^{i \t} e^{iJ}) + \I (e^{i \t} \O ) \right] ~,
}
for $||\chi_1||^2 = ||\chi_2||^2 = e^A$. As stated earlier, we will study
the case of a strict $SU(3)$-structure for which $\theta = 0$ and hence
\eq{\label{su3poly}
\psi_+ &= \frac{1}{8} e^A \left[  \I(e^{iJ}) + v\wedge \R (\O) \right]~, \\
\psi_- &=   \frac{1}{8} e^A \left[ v \wedge \R(e^{iJ}) + \I (\O) \right] ~.
}
Substituting the above expressions in the supersymmetry equations \eqref{SUSY}, we will
derive the restrictions on the intrinsic torsion of the $SU(3)$-structure imposed by supersymmetry.

\section{A class of solutions to the supersymmetry equations}\label{classification}
In this section, we derive a class of solutions to the supersymmetry equations \eqref{SUSY} by inserting the strict $SU(3)$-structure polyforms \eqref{su3poly}.

The first constraint \eqref{SUSYa} yields
\begin{subequations}\label{IIBNSNS}
\begin{align}
d \left( e^{2A - \phi} J \right) &= 0~,   \\
d \left( e^{2A - \phi} v \wedge \R \O \right) - e^{2A-\phi} H \wedge J &= 0~, \\
d \left(e^{2A - \phi} J \wedge J \wedge J \right) + 3! \, e^{2A- \phi} H \wedge v \wedge \R \O &= 0~.
\end{align}
\end{subequations}
These in turn determine
\eq{
0 &= W_1 = W_3 = V_2 = T_2~, \\
2dA-d\phi &= - W_4 - \frac{2}{3}\R E \, v~.
}
Upon decomposing the NSNS field-strength $H$ with respect to the
$SU(3)$-structure as
\eq{\label{Hdecomp}
H &= H^R \R\O + H^I \I\O + \left( H^{(1,0)} + H^{(0,1)} \right)\wedge J + H^{(2,1)} + H^{(1,2)} \\
&+ v \wedge  \left( H_v^{(1,1)}  + H^0_v J + H^{(0,1)}_v \lrcorner \O + H^{(1,0)}_v \lrcorner \overline{\O} \right)~,
}
where $H^{(2,1)}$ and $H_v^{(1,1)}$ are primitive, we also find expressions for several of the components in terms of torsion classes from
\eqref{IIBNSNS}.  Using \eqref{id3}, we find:
\eq{\label{IIBH}
H^I = - \frac{1}{3} \R E~,& \quad
H^{(1,0)} = V_1~, \quad \\
H_v^{(1,1)} = - \R W_2~, \quad
H_v^0 = 0~&, \quad
H_v^{(1,0)} = \frac{1}{2i}\left(W_4^{(1,0)} + W_0^{(1,0)} -  W_5\right)
~.
}

The exterior derivatives of the the $SU(3)$-structure tensors now read
\begin{subequations}
\label{IIBSU(3)}
\begin{align}
dv &= R J + \R (\overline{V_1} \lrcorner \O) + T_1 + v \wedge W_0~,  \\
dJ &= -d(2A-\phi) \wedge J~, \\
d \O &= W_2 \wedge J + (\overline{W_5} + E \, v) \wedge \O
+ v \wedge S~.
\end{align}
\end{subequations}
We define a rescaled metric $g_{M_7} = e^{-2A + \phi} \check{g}_{M_7}$ and rescale the $SU(3)$-structure tensors accordingly as $\{v,\, J,\, \Omega\} = \{e^{-A+\phi/2}\check{v},\, e^{-2A + \phi}\check{J},\,  e^{-3A + 3\phi/2}\check{\Omega}\}$ to obtain
\sub{
d \check{v} &= \check{R} \check{J} + \R (\overline{\check{V}}_1 \lrcorner \check{\O}) + \check{T}_1 + \check{v} \wedge \check{W}_0~, \label{vcheck} \\
d \check{J} &= 0~, \label{Jcheck} \\
d \check{\O} &= \check{W}_2 \wedge \check{J} + (\overline{\check{W}}_5 + i \I \check{E} \, \check{v}) \wedge \check{\O}
+ \check{v} \wedge \check{S}~, \label{Omegacheck}
}
where
\eq{
\check{W}_0 = W_0 + \frac{1}{2} W_4~, \qquad
\check{W}_5 = W_5 - \frac{3}{2} W_4~,
}
and
\eq{
\begin{alignedat}{5}
R    &{}= e^{A - \frac\phi 2} \check{R}~, \quad &
\I E &{}= e^{A - \frac\phi 2} \I \check{E}~, \quad&
V_1  &{}= \check{V}_1~,
\\
W_2 &{}= e^{-A+\frac{\phi}{2}} \check{W}_2~, \quad &
T_1 &{}= e^{-A + \frac\phi 2} \check{T}_1 ~, \quad &
S   &{}= e^{-2A + \phi} \check{S}~.
\end{alignedat}}
We note that the condition $d\check{J} = 0$ means that the six-dimensional leaves $M_6$ transverse to $\check{v}$
admit a symplectic structure.

Turning to the second constraint \eqref{SUSYb} we obtain:
\begin{subequations}
\label{IIBRRstar}
\begin{align}
e^{3A} \star_7  F_7 &= 0~, \\
e^{3A} \star_7  F_5 &= d \left( e^{3 A - \phi} v \right) + 2 \m e^{2A-\phi} J~,  \label{starF5} \\
- e^{3A} \star_7 F_3 &= d \left(e^{3 A - \phi} \I \O \right) - e^{3A-\phi} H \wedge v + 2 \m  e^{2 A - \phi} v \wedge \R \O~,  \\
e^{3A} \star_7  F_1 &= -\frac{1}{2} d \left( e^{3A - \phi} v \wedge J \wedge J \right) -  e^{3A - \phi} H \wedge \I \O - \frac{1}{3} \m  e^{2A - \phi} J \wedge J \wedge J~.
\end{align}
\end{subequations}
From these equations, employing \eqref{IIBH} and \eqref{IIBSU(3)} and the set of identities \eqref{id1}, \eqref{id2}, we can
obtain expressions for the magnetic RR fluxes $F_p$, $p=1,3,5,7$. We give these in \eqref{IIBRR} in the summary below.

Finally, the third constraint \eqref{SUSYc} reads
\begin{equation}
F_5 \wedge J - F_3 \wedge v \wedge \R \O  - \frac{1}{3!} F_1 \wedge J \wedge J \wedge J= 4 \m e^{-A-\phi} \vol_7~,
\end{equation}
and plugging in the expressions for the RR fields we conclude that
\eq{
3R + 6 m e^{-A} + 4 H^R + 2 \I E = 0 ~.
}

\subsection{Summary}

Let us summarize our results. The differential constraints imposed on the $SU(3)$-structure by supersymmetry are:
\begin{subequations}
\begin{align}
dv &= R J + \R (\overline{V_1} \lrcorner \O) + T_1 + v \wedge W_0~,  \\
dJ &= -d(2A-\phi) \wedge J~, \\
d \O &= W_2 \wedge J + (\overline{W_5} + E \, v) \wedge \O
+ v \wedge S~.
\end{align}
\end{subequations}
The expression for the NSNS field is:
\eq{\label{Hdecomp}
H &= -\frac14\left(3R + 6me^{-A} + 2 \I E \right) \R\O - \frac{1}{3} \R E ~\I\O + 2 \R V_1 \wedge J + 2 \R(H^{(2,1)}) \\
&+ v \wedge  \left( -\R W_2  + \I\left((W_4^{(1,0)} + W_0^{(1,0)} -  W_5) \lrcorner \overline{\O}\right) \right)~.
}
The expressions for the RR fields are:
\begin{subequations}\label{IIBRR}
\begin{align}
e^\phi F_1 &= \left(2\I E + 4 m e^{-A}\right) v + 2\I (X_1^{(1,0)}) ~,\\
e^\phi F_3 &= \frac14 \left(-2 m e^{-A} + 3 R - 2 \I E \right) \I \O - 2 \I V_1 \wedge J + v \wedge \I W_2  \nn \\
&+ 2 \I(H^{(2,1)}) -  \R S + X_3 \lrcorner(v \wedge \R \O)~, \\
e^\phi F_5 &= \frac12  \left( R + 2m e^{-A}\right) v  \wedge J \wedge J - \I ( X_5^{(1,0)}) \wedge J \wedge J  \nn \\
&- v \wedge J \wedge T_1 + 2 v \wedge  \R V_1  \wedge \I \O~,  \\
e^\phi F_7 &= 0~,
\end{align}
\end{subequations}
with
\eq{
X_1 &\equiv dA + W_0 + 3 W_4 - 2(W_5 + \overline{W_5})~, \\
X_3 &\equiv dA - W_4 + W_5 + \overline{W_5} ~, \\
X_5 &\equiv dA - W_0 - W_4 ~.
}

The above solution to the supersymmetry equations also solves the equations of motion if and only if the Bianchi identities for the
NSNS and RR fields are imposed in addition.

\section{A new class of solutions}
\label{newclass}

We make the following Ansatz:
\begin{equation}
H + i e^\phi F_3 = 2 H^{(2,1)}~,
\end{equation}
and recall that $H^{(2,1)}$ is primitive. This leads to $v \lrcorner dA = 0 = v \lrcorner d\phi$ and the following restrictions on the torsion classes:
\eq{
0 = \R E = V_1 = W_2 &= S = W_5 - W_0^{(1,0)} - W_4^{(1,0)}~, \\
\I E = -2me^{-A}~, \quad
R = -\frac{2}{3}me^{-A}~&, \quad
W_0 = - dA~, \quad
W_4 = - 2dA+d\phi~.
}
We thus have
\begin{subequations}
\begin{align}
dv &= -\frac{2}{3} m e^{-A} J + T_1 - v \wedge dA ~,  \\
dJ &= -d(2A-\phi) \wedge J~, \\
d \O &= \left(-3dA+d\phi - 2i me^{-A} v\right) \wedge \O~,
\end{align}
\end{subequations}
or in terms of the rescaled $SU(3)$-structure
\begin{subequations}
\begin{align}
d\check{v} &= -\frac{2}{3} m e^{-2A + \phi/2} \check{J} + \check{T}_1
- \check{v} \wedge \left(2dA - \frac12d\phi\right)~, \\
d \check{J} &= 0~, \\
d \check{\O} &= \left(-\frac{1}{2} d\phi -  2i m e^{-2A+\phi/2} \check{v}\right) \wedge \check{\O}~.
\end{align}
\end{subequations}
From the differential equations for $\{\check{J}\,,\check{\O}\}$
we conclude that $\check{M}_6$ is K\"{a}hler. In what follows we will introduce the exterior derivative $d_6$ on $\check{M}_6$, Dolbeault operators $\partial$, $\bar\partial$ so that $d_6 = \partial + \bar\partial$, and $d_6^c = i (\bar{\p} - \p )$.
The remaining RR fields read
\eq{
F_1 &= - d_6^c e^{-\phi}~,\\
F_5 &= \frac{2}{3} m e^{-6A + 3\phi/2} \check{v} \wedge \check{J}^2
+ \frac12 d_6^c(e^{-4A + \phi})  \wedge \check{J} \wedge \check{J}
- e^{-4A + \phi} \check{v} \wedge \check{J} \wedge \check{T}_1~.
}
Let us now examine the Bianchi identities.
The first Bianchi identity, $dF_1 = 0$, enforces
\eq{
\p \bar\p  e^{-\phi}=  0~,
}
which is solved by setting $\phi = -\log( \varphi + \overline{\varphi})$, with $\varphi$ holomorphic.
Next, the three-form Bianchi identities $d H = 0$ and $d F_3 - H \wedge F_1 = 0$ yield the constraints
\eq{\label{constraint}
\partial H^{(1,2)} + \bar{\p} H^{(2,1)} = \bar{\p} H^{(1,2)} =  
\partial H^{(1,2)}
- \partial \phi \wedge H^{(1,2)}
+ \bar\partial \phi \wedge H^{(2,1)} = 0~.
}

In analyzing the Bianchi identity for $F_5$, we will invoke the
results of \cite{Couzens:2017nnr}. The authors of \cite{Couzens:2017nnr}
study supersymmetric solutions which descend from the solutions analyzed here, upon setting $H^{(2,1)} = 0$.
However, even when $H^{(2,1)} \neq 0$, the $SU(3)$-structure of \cite{Couzens:2017nnr} and the expressions for $F_1$ and $F_5$ can be identified with the ones presented in this section. The map identifying the tensors there (left-hand side),
with the tensors here (right-hand side) is
\eq{
\begin{alignedat}{5}
P &{}= \p \phi~,                   \qquad & Q &{}= -\frac12 d_6^c \phi~,     \qquad & F^{(2)} &{}= -e^{3A} \star_7 F_5~,\\
\Delta &{}= A - \frac{1}{4}\phi ~, \qquad & e^{2\Delta} K &{}= \check{v}~, \qquad & \widetilde{g}_6 &{}= m^2 \check{g}_6~,
\end{alignedat}
}
and in particular, \eqref{starF5} is identified with (2.58) of \cite{Couzens:2017nnr}.\footnote{One has to take into account that
we work in the string frame whereas the Einstein frame is used in \cite{Couzens:2017nnr}. In addition, we use a different orientation on AdS$_3$.}
The authors of \cite{Couzens:2017nnr} showed that the Bianchi identity
for $F_5$, $dF_5 = 0$, amounts to
\eq{
\nabla^2 ( R - 2 |\p \phi|^2 ) - \frac12 R^2 + R_{ij} R^{ij} + 2 |\p \phi|^2 R - 4 R_{ij} \p^i \phi \overline{\p}^j \phi  = 0~,
}
which they refer to as the ``master equation''. In the above, $R$ and $R_{ij}$
are respectively the Ricci scalar and the Ricci tensor of $\check{g}_6$, and  contractions are also made using $\check{g}_6$. This master equation generalizes the one derived in \cite{Kim:2005ez} by including a varying axio-dilaton.
For the case at hand the Bianchi identity
of $F_5$ is $dF_5 = H \wedge F_3$, and the term on the right-hand side (a ``transgression'' term) modifies the master equation, which now becomes:
\eq{\label{master}
\nabla^2 ( R - 2 |\p \phi|^2 ) - \frac12 R^2 + R_{ij} R^{ij} + 2 |\p \phi|^2 R - 4 R_{ij} \p^i \phi \overline{\p}^j \phi - \frac{8}{3} e^{-\phi} H^{(2,1)}_{ijk} {(H^{(1,2)})}^{ijk} = 0~.
}

As noted above, in the limit $H^{(2,1)} = 0$ the present class of
solutions and master equation \eqref{master} reduce to the ones
of \cite{Couzens:2017nnr}. Further setting the axio-dilaton
to zero, they reduce to the ones studied in \cite{Kim:2005ez}. Starting with the latter, the authors of \cite{Donos:2008ug} ``turned on'' a three-form flux $G = G^{(1,2)}$, and taking the limit of vanishing axio-dilaton we recover their results. See Figure \ref{classes}. 

Finally, the supersymmetry preserved by the class of solutions in this section enhances to $\mathcal{N}=2$, and the dual field theories are $(0,2)$ SCFTs \cite{Couzens:2019iog}.
The vector field dual to $v$ generates a $U(1)$ symmetry of the solutions, corresponding to the R-symmetry of the $(0,2)$ SCFTs.
Thus we expect that a geometric dual of $c$-extremization exists 
for this class of solutions and would be very interesting to identify it.

\vskip 1cm

\noindent
{\bf Acknowledgements}

\noindent
We would like to thank N.~Macpherson and A.~Tomasiello for useful discussions. The work of A.P. is supported by is supported by Agence Nationale de la Recherche LabEx grant ENS-ICFP ANR-10-LABX-0010/ANR-10-IDEX-0001-02 PSL. D.P. has been financially supported in part by INFN and by the ERC Starting Grant 637844-HBQFTNCER

\appendix

\section{Conventions \& identities}

\subsection*{Clifford algebra decomposition}
The ten-dimensional gamma matrices are decomposed as follows 
\begin{equation}
\Gamma_\mu = e^A \gamma_\mu^{(3)} \otimes \mathbb{I} \otimes \sigma_3~,\qquad
\Gamma_m = \mathbb{I} \otimes \gamma_m \otimes \sigma_1~,
\end{equation}
where $\gamma_\mu^{(3)}$ span Cliff$(1,2)$, $\gamma_m$
span Cliff$(7)$ and the indices are spacetime indices.

We take $\gamma_\mu^{(3)}$ to be real, and $\gamma_m$ imaginary and antisymmetric. In particular we have 
\begin{equation}
\gamma_m = \gamma_m^\dagger~,\qquad
\gamma_m = - \gamma_m^t~, \qquad
\gamma_m = - \gamma_m^*
\end{equation}
\begin{equation}
\gamma^{-1}_0 \gamma_\mu^{(3)} \gamma_0 = - (\gamma_\mu^{{(3)}})^\dagger~, \qquad
\gamma^{-1}_0 \gamma_\mu^{(3)} \gamma_0 = - (\gamma_\mu^{{(3)}})^t~,
\qquad
\gamma_\mu^{{(3)}} = (\gamma_\mu^{{(3)}})^*~.
\end{equation}

It follows that the ten-dimensional intertwiners
$A$, $C$, and $D$ defined by
\begin{equation}
A^{-1} \Gamma_M A = \Gamma_M^\dagger~, \qquad 
C^{-1} \Gamma_M C = - \Gamma_M^t~, \qquad 
D^{-1} \Gamma_M D = \Gamma_M^*~, 
\end{equation}
are
\begin{equation}
A = \gamma_0 \otimes \mathbb{I} \otimes \sigma_1~, \qquad 
C = \gamma_0 \otimes \mathbb{I} \otimes \mathbb{I}~, \qquad
D = \mathbb{I} \otimes \mathbb{I} \otimes \sigma_3~.
\end{equation}

\subsection*{Spin$(7)$}
As noted above, we work with the Majorana representation of Cliff$(7)$,
for which the gamma matrices are imaginary and antisymmetric. The charge-conjugate of a Spin$(7)$ spinor
is the complex conjugate, and a Majorana spinor is real.
The basis elements of $\mathrm{Cliff(7)}$ are related via the identity
\begin{equation}
\gamma_{m_1 \dots m_k} = \frac{i}{(7-k)!} (-1)^{k(k-1)/2} \,
\epsilon_{m_1 \dots m_k m_{k+1} \dots m_7} \gamma^{m_{k+1} \dots m_7} ~.
\end{equation}

As discussed in section \ref{gstructures}, a pair of nowhere-vanishing Spin$(7)$ Majorana spinors $\chi_1$ and $\chi_2$ define an $SU(3)$-structure $\{v,\, J,\, \Omega\}$ in seven dimensions.
For the strict $SU(3)$-structure $(\t = 0$, or equivalently, $\chi_1^t \chi_2 = 0)$, we introduce a Dirac spinor $\eta$ as
\eq{\label{etadef}
\chi_1 = \frac{1}{\sqrt{2}} \left( \eta + \eta^* \right)~, \qquad
\chi_2 = \frac{i}{\sqrt{2}} \left( \eta^* - \eta \right)~.
}
The bilinears that can be constructed using $\eta$ are $\eta^\dagger \g_{m_1 \dots m_k} \eta$ and $\eta^t \g_{m_1 \dots m_k} \eta$.
In terms of the $SU(3)$-structure, $\eta$ satisfies
\eq{\label{bilinears}
\begin{alignedat}{6}
\eta^\dagger \eta      &= e^A ~,                       & \qquad \qquad \eta^t         \eta &{}={} 0~, \\
\eta^\dagger \g_m \eta &= e^A v_m~,                   & \qquad \qquad \eta^t \g_m    \eta &{}={} 0~, \\
\eta^\dagger \g_{mn} \eta &= i e^A J_{mn}~,            & \qquad \qquad \eta^t \g_{mn} \eta &{}={} 0~, \\
\eta^\dagger \g_{mnp} \eta &= 3 i e^A v_{[m} J_{np]}~, & \qquad \qquad \eta^t \g_{mnp}\eta &{}={} -i e^A \O_{mnp}~.
\end{alignedat}
}
These can then be used to deduce the expressions \eqref{su3poly} for the polyforms.
In the case of the $G_2$-structure $(\t = \pi/2$, or $\chi_1 = \chi_2)$, we may instead consider $\chi_1 = \chi_2 = \frac{1}{\sqrt{2}}(\eta + \eta^*)$, where $\eta$ satisfies the above equations.

\subsection*{Identities}
The $SU(3)$-structure is normalized as follows:
\eq{
\O_{mpq} \overline{\O}^{npq} &= 2^3 \left(\delta_m{}^n - i J_m^{\phantom{m}n} - v_m v^n\right)~, \quad
\O_{mnp} \overline{\O}^{mnp} = 3! 2^3~, \\
\e_{m_1...m_7} &= \frac{7!}{3!2^3} v_{[m_1} J_{m_2 m_3} J_{m_4 m_5} J_{m_6 m_7]}~.
}
Given the above normalization, we derive a number of identities necessary to obtain the RR fields $F_{1,3,5}$ from
their Hodge duals $\star_7 F_{1,3,5}$. Duals of the $SU(3)$-structure are given by
\eq{\label{id1}
\star_7 J = \frac12 v \wedge J \wedge J~, \quad
\star_7 (v \wedge J) = \frac12 J \wedge J~, \quad
\star_7 \O = i v \wedge \O~.
}
Duals for arbitrary primitive $(p,q)$-forms $\o^{(p,q)}$ are given by
\eq{\label{id2}
\star_7 (\o^{(1,0)} \wedge J) = i v \wedge \o^{(1,0)} \wedge J
~,& \quad
\star_7 (\o^{(1,0)} \wedge J \wedge J) = 2 i v \wedge \o^{(1,0)}
~,\\
\star_7 (v \wedge (\o^{(0,1)} \lrcorner \O )) =  - i \o^{(0,1)} \wedge \O
~,& \quad
\star_7 (\o^{(0,1)} \lrcorner \O ) =  - i v \wedge \o^{(0,1)} \wedge \O ~,\\
\star_7 \o^{(1,1)} = - v \wedge J \wedge \o^{(1,1)}
~,& \quad
\star_7 (\o^{(1,1)} \wedge J) = - v \wedge \o^{(1,1)}~, \\
\star_7 \o^{(2,1)} = - i v\wedge \o^{(2,1)}
~.
}
We also make use of the identities
\eq{\label{id3}
( \o^{(0,1)} \lrcorner \O ) \wedge J = - i \o^{(0,1)} \wedge \O
~, \quad
( \o^{(0,1)} \lrcorner \O ) \wedge \overline{\O} = 4 \o^{(0,1)} \wedge J \wedge J~,
}
in order to obtain the components of $H$.

\bibliography{3}
\bibliographystyle{utphys}

\end{document}